# Widely color-temperature low-luminosity-loss electrochromic-tuned white light-emitting diodes

Yu-Yi Kuo,[1] Chiu-Chang Huang,[1] Wei-Ting Chen,[1] Ting-Hsiang Chang,[2] Hsin-Che Lu,[2] Kuo-Chuan Ho,[2,*] and Chih-Yu Chao[1,3,*]

[1]Department of Physics, National Taiwan University, Taipei 10617, Taiwan
[2]Department of Chemical Engineering, National Taiwan University, Taipei 10617, Taiwan
[3]Biomedical & Molecular Imaging Center, National Taiwan University College of Medicine, Taipei 10051, Taiwan
*kcho@ntu.edu.tw
*cychao@phys.ntu.edu.tw

**Abstract:** Light-emitting diodes (LEDs) are efficient light sources extensively applied in people's daily life nowadays, with profound effect on their psychological and physiological state subtly and deeply. The correlated color temperature (CCT) the eyes perceive is a key parameter, but the CCT tuning technology of LEDs remains in the development stage. In this study, an electrochromic (EC) material is employed in the tuning device, exhibiting a wide CCT tuning range from 3,200 K to 6,900 K, varying almost along the black-body locus. The driving voltage is lower than 2.4 V, with luminosity loss less than 16%. Using the EC materials, the performance of CCT tuning technology is able to be further enhanced.

## 1.  Introduction

Since the publication of the first report on the technology [1], light-emitting diode (LED) has been widely studied for more than 50 years. The invention of high-power InGaN LEDs led to the realization of white LEDs technology [2], which has become a prevalent illumination technology nowadays. There are two common methods for the fabrication of white LEDs. The first and the most direct one is blending the lights of red, green, and blue (RGB) LEDs, which is the so-called RGB system [3]. The other is the integration of yellow or green-and-red-mixed phosphor with high-intensity blue LEDs, producing a broad spectrum perceived as white light to human eyes [4], which is called phosphor-converted white LEDs.

There has been extensive study on the impact of artificial light on humans [5-8]. It has been applied in therapeutic treatments for sleeping problem [9], depression [10], and resetting of the circadian clock [11], among others. However, adverse physiological effects have often been reported, such as disruption of the circadian clock [12], interrupting the biological processes dependent on the cycle of day and night [13]. Moreover, it has been proven that human performances, in both physiological and psychological aspects, are closely associated with their perception of ambient illumination, especially the CCT [14]. However, there is still no simple way available for controlling the CCT of white LEDs precisely. Therefore, the development of CCT-tunable white LEDs becomes critical in this era [15].

In our previous work, we introduced a method based on a liquid crystal (LC) cell for manipulating the CCT of white LEDs electrically, but with noticeable loss of luminosity due to the polarizers in the LC cell system [16]. The drawback was improved by adding dichroic dye to the LC cell, which utilizes the LC molecules to align the doped dye in different directions electrically [15]. Nonetheless, the variation range of CCT is shorter due to the concentration limitation of dye. Despite the discrepancy in results, both of the tuning systems absorbed abundant intrinsic blue light emitted by LEDs via the tuning device, suggesting a possible way to improve the performance of the CCT tuning device by finding a material with more specific absorption spectrum on blue light of the LEDs. A promising candidate is electrochromic (EC) material, which is able to alter color performance when electrons transfer [17], and as a result the absorbance curve of some is deformed in visible spectrum, decreasing the penetration in certain wavelength [18]. This property matches the CCT tuning mechanism we require, so long as a suitable EC material, with absorption spectrum peaking in the region 460~480 nm after being electrically manipulated, is applied. And the most important point is that the manipulation of EC material requires no polarizer, which would largely decrease the unnecessary luminosity loss as in the LC cell system.

In this study, 4,4'-biphenyl dicarboxylic acid diethyl ester (PCE), an EC material, was used in the fabrication of the EC cell [19]. After being electrically driven, there was an absorption peak at around 460 nm, matching the goal of absorbing blue light in white LEDs. In our experiments, the EC device tuned the CCT of the LED to a range from about 3,200 K to 6,900 K, almost in line with the black body locus (BBL) curve, demonstrating the cold white, neutral white, and warm white light in a bulb. Furthermore, the absence of polarizers reserves 84% luminosity while exhibiting a very wide CCT tuning range. This property fulfills the goal of an ideal CCT tunable device.

## 2. Experiments

### 2.1. Preparation of pc-WLEDs

Surface-mount device (SMD) LEDs were chosen in our system. A 3 × 3 SMD LEDs array was welded on a printed circuit board, with one royal blue LED (Luxeon Rebel, Philips) located at the center and 8 blue LEDs (Luxeon Rebel, Philips) in the peripheral area. The yellow phosphor (LWB GmbH) and the red phosphor ($Ca_2Si_5N_8:Eu^{2+}$, Dott Technology) were mixed in various ratios for testing to realize the best performance. To spread the phosphor uniformly, the mixture was blended in 5.5 ml silicone gel (Sylgard 184, Dow Corning) and then spin-coated on a glass substrate.

### 2.2. Preparation of EC cells

PCE solution was prepared by the modified method from the previous study [19]. Briefly, 25 mM of 4,4'-biphenyl dicarboxylic acid diethyl ester (Tokyo Chemical Industry) and 50 mM of lithium perchlorate (Sigma-Aldrich) were dissolved in N-methyl-2-pyrrolidinone (Sigma-Aldrich). The empty EC cell was fabricated as follows: two glass pieces with indium tin oxide coating (ITO glasses, ~7 Ω/□) were cut into suitable size to match the phosphor substrate. ITO glasses were cleaned and dried completely before use. Spacer films were sliced, placed by the edge of the ITO glasses, sandwiched between the ITO coating sides, and heated to be melted. After cooling down, PCE was injected into the cell gap with the aid of capillary effect, and all sides were then immediately sealed with UV glue when fulfilled. The configuration is illustrated in Fig. 1(a).

### 2.3. Combination of pc-WLEDs and EC cells

The SMD LEDs were set on the base of a processed LED bulb. Both of the substrate and the EC cell were held above the SMD LEDs by a hollow copper bulk, which facilitated dissipating the excess heat on the EC cell. Fig. 1(b) illustrates the assembly. A dome-shaped diffuser was set on the top to uniformize the CCT performance, which has been demonstrated in our previous research [16].

## 3. Results and Discussions

Employing the parameters of our previous work [16], the 1931 Commission Internationale de L'éclairage (CIE 1931) chromaticity coordinates of the assembly was measured by a spectrometer (SD1200, OTO Photonics) and shown in Fig. 2(a), with the driving voltage of the EC cell from 0 to 2.3 V. The phosphor substrate was prepared with 3,000 mg yellow phosphor, and the LEDs array consisted of 9 royal blue LEDs. Before the driving voltage applied, the CCT coordinate of the mixed light was located outside the binning system of white LEDs. After the driving voltage applied, it ends to about 5,400 K. The tuning range is only about 2,000 K in the cold white region, smaller than the best performance in our previous work but better than the guest-host LC system [15].

The location of the starting point in Fig. 2(a) deviates from the BBL, indicating the combination of LEDs arrangement and the phosphor concentration requires further optimization. Fig. 2(b) shows the CIE coordinates of two bare LEDs arrays. The CIE coordinate of 9 royal blue array is lower than the 8 blue and 1 royal blue LEDs array. Furthermore, the connection segment to 6,500 K of the 8 blue and 1 royal blue LEDs array matches the tangential slope over the BBL curve more than the 9 royal blue LEDs array. Therefore, the 8 blue and 1 royal blue arrangement is a better LEDs array for our experiment. To blend the light more uniformly, the dome-shaped diffuser is installed, and the uniformity before and after the installation is shown in Fig. 3(a). Fig. 3(b) shows the variation of CCT of the 8 blue and 1 royal blue LEDs array, with the optimized phosphor ratio 200 mg yellow phosphor and 600 mg red phosphor. The location of the starting point is nearer to the BBL, with wider CCT variation ranging from about 3,200 K to 6,900 K. Besides, in the region between 4,000 K to 5,700 K, the variation track moves almost alongside the BBL. It shows a novel potential for the artificial light to simulate distinct natural light sources with the aid of EC system besides the cholesteric films [20]

To further measure the intensity of the visible spectra and the luminosity of whole LED bulb before and after the EC cell is driven, a photometric integrating sphere (acquired by AMA Optoelectronics Inc.) is used and the results are shown in Fig. 4. The spectra remain nearly unchanged under 2.0 V or less driving voltage applied. The peak around 480 nm decreased when the driving voltage excessed 2.0 V, as expected. Though the blue light of the whole bulb decreased significantly when the EC cell was driven, its luminosity was not affected very much. Table 1 shows the luminous flux measured by the photometric integrating sphere with respect to different driving voltage on the EC cell. Before the EC cell was driven, the luminous flux stood at about 130.86 lm, which dropped to about 110.03 lm, for a difference of 15.9%, after the driving voltage rose to 2.4 V. By contrast, the LC cells applied in our previous work [16], equipped with patterned polarizers, absorbed 90% of the light in the working region, leading to significant reduction of the brightness of LEDs. Therefore, the low consumption of light suggests that EC cells may pose as a better option in CCT tuning method.

## 4. Conclusion

With the white LED being a potential candidate for the next generation illumination, a smart solution to the CCT fluctuation is required. This paper proposes a polarizer-free, EC material-based CCT-tunable device, capable of tuning white light in a wide range, boasting the merit of low luminosity loss (only 15.9% in luminous flux), which is much smaller than that in the LC-based tuning system (larger than 50 %). Moreover, the CCT of the modulated light based on EC system ranges from 3,200 K to 6,900 K, covering the performance of the LC- and dichroic dye-based tuning systems. This tuning method can be operated easily and tune the CCT precisely without complex control circuit. With the aid of EC technologies, CCT-tunable white LEDs can be further improved.

## Declaration of interests

The authors declare that they have no known competing financial interests or personal relationships that could have appeared to influence the work reported in this paper.

## Acknowledgments

The author CYC would like to acknowledge the support from National Taiwan University, the Ministry of Science and Technology (MOST 105-2112-M-002-006-MY3), and the Ministry of Education (MOE 106R880708) of the Republic of China. The author KCH also would like to thank the support from the Ministry of Science and Technology (MOST 106-2221-E-002-180-MY3).

**[Color should not be used for any figures in print]**

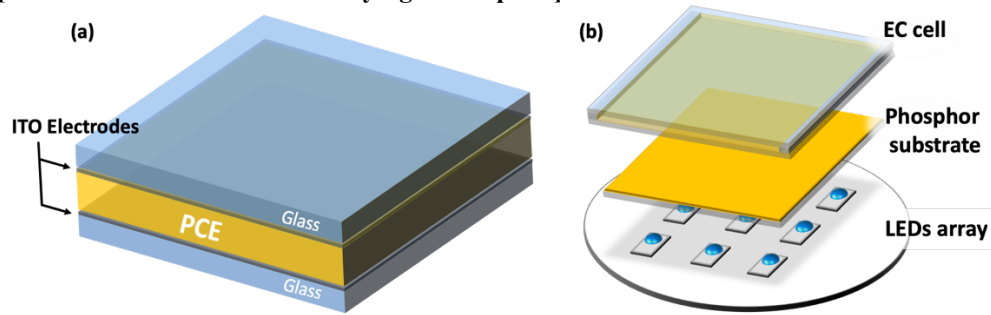

Fig. 1. (a) The schematic structure of the EC cell. (b) The arrangement of LEDs array, phosphor substrate and EC cell.

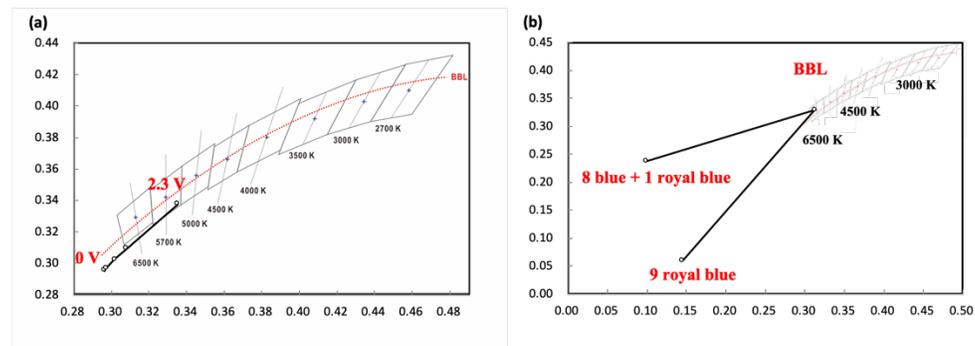

Fig. 2. (a) The variation of CIE coordinates of the assembly consisting of the LEDs array (9 royal blue LEDs), the phosphor substrate (3,000 mg yellow phosphor in 5.5 ml silicone gel). (b) The CIE coordinates of two different LEDs arrays without phosphor substrate.

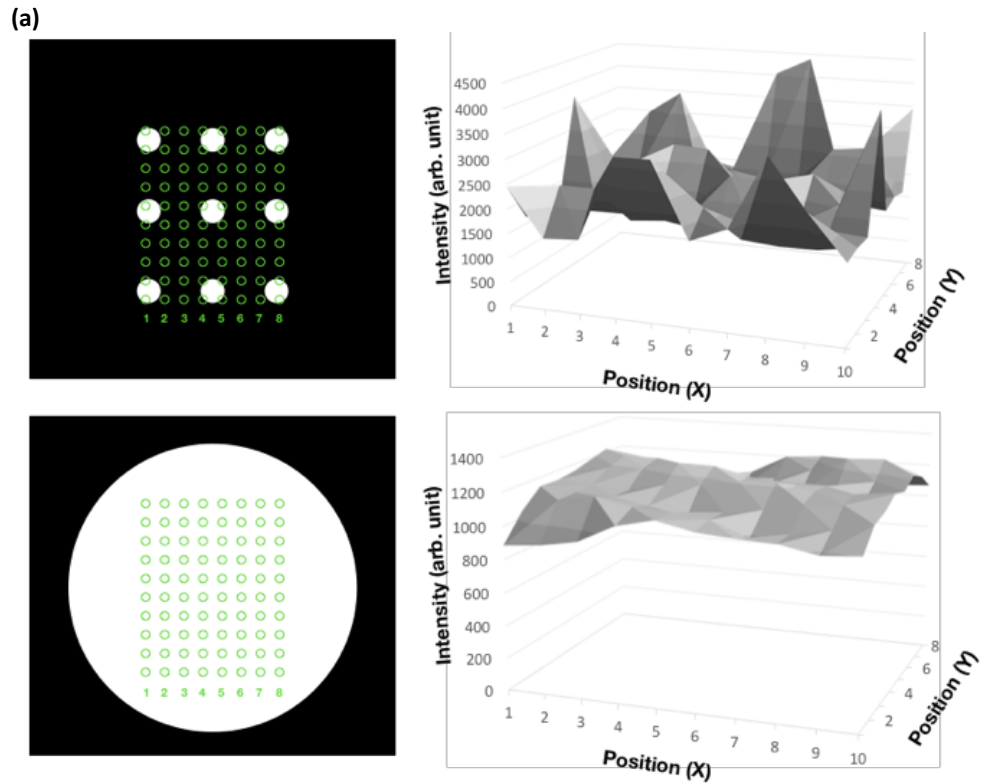
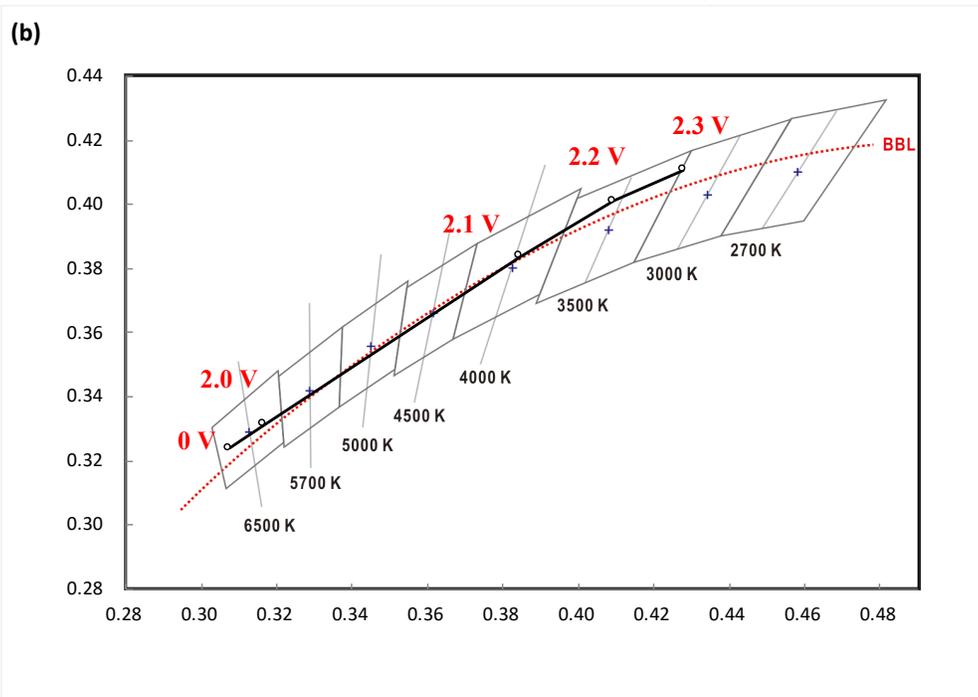

Fig. 3. (a) The uniformity of intensity measurements of 80 points without (middle) and with (right) the dome-shaped diffuser. (b) The variation of CIE coordinates measured by spectrometer (SD1200, OTO Photonics) on LEDs array (8

blue and 1 royal blue LEDs) with the optimized phosphor substrate (200 mg yellow and 600 mg red in 5.5 ml silicone gel).

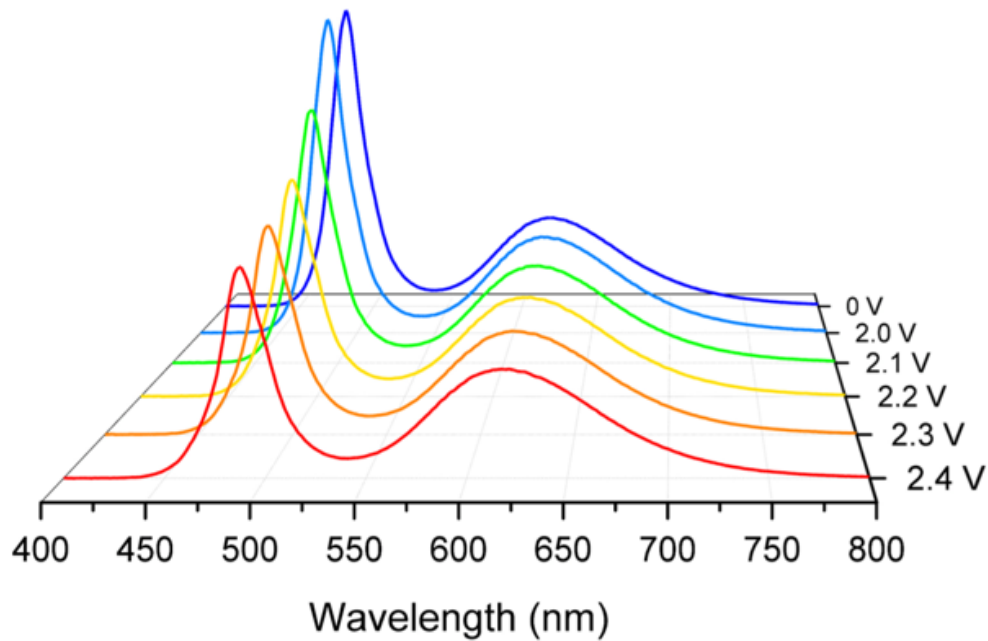

Fig. 4. The spectra measured by the photometric integrating sphere under different driving voltages.

Table 1. Integrated luminous flux with respect to the driving voltage applied on the EC cell

| Voltage (V) | 0 | 2.0 | 2.1 | 2.2 | 2.3 | 2.4 |
|---|---|---|---|---|---|---|
| Flux (lm) | 130.86 | 131.32 | 122.76 | 115.64 | 112.10 | 110.03 |